\documentclass{aastex62}

\def\qqquad{\hskip3em\relax}

\graphicspath{{./}{figures/}}

\usepackage{lineno}

\usepackage{xspace}
\usepackage{amsmath}

\usepackage{savesym}
\savesymbol{tablenum}
\usepackage{siunitx}
\restoresymbol{SIX}{tablenum}

\newcommand{\Abacus}{\textsc{Abacus}\xspace}
\newcommand{\singlestep}{\texttt{singlestep}\xspace}
\newcommand{\convolution}{\texttt{convolution}\xspace}
\newcommand{\hal}{\texttt{hal}\xspace}

\newcommand{\pkdgrav}{\textsc{Pkdgrav3}\xspace}
\newcommand{\gadget}{\textsc{Gadget3}\xspace}
\newcommand{\ramses}{\textsc{Ramses}\xspace}

\newcommand{\kny}{\ensuremath{k_\mathrm{Nyquist}}}

\DeclareSIUnit \hMpc {\ensuremath{\mathit{h}^{-1}} \mathrm{Mpc}}
\DeclareSIUnit \hkpc {\ensuremath{\mathit{h}^{-1}} \mathrm{kpc}}
\DeclareSIUnit \perhMpc {\ensuremath{\mathit{h} \mathrm{Mpc}^{-1}}}
\DeclareSIUnit \hMsun {\ensuremath{\mathit{h}^{-1}} M_\odot}

\newcommand{\bfF}{\mathbf{F}}
\newcommand{\bfr}{\mathbf{r}}

\shorttitle{\Abacus Code Comparison}
\shortauthors{Garrison et al.}

\defcitealias{S2016}{S2016}

\begin{document}

\title{A High-Fidelity Realization of the Euclid Code Comparison $N$-body Simulation with \textsc{Abacus}}

\correspondingauthor{Lehman Garrison}
\email{lgarrison@cfa.harvard.edu}

\author{Lehman H.~Garrison}
\affil{Harvard-Smithsonian Center for Astrophysics, 60 Garden St., Cambridge, MA 02138}

\author{Daniel J.~Eisenstein}
\affil{Harvard-Smithsonian Center for Astrophysics, 60 Garden St., Cambridge, MA 02138}

\author{Philip A.~Pinto}
\affil{Steward Observatory, University of Arizona, 933 N. Cherry Ave., Tucson, AZ 85121}

\begin{abstract}
We present a high-fidelity realization of the cosmological $N$-body simulation from the \cite{S2016} code comparison project.  The simulation was performed with our \textsc{Abacus} $N$-body code, which offers high force accuracy, high performance, and minimal particle integration errors.  The simulation consists of $2048^3$ particles in a $500\ h^{-1}\mathrm{Mpc}$ box, for a particle mass of $1.2\times 10^9\ h^{-1}\mathrm{M}_\odot$ with $10\ h^{-1}\mathrm{kpc}$ spline softening.  \textsc{Abacus} executed 1052 global time steps to $z=0$ in 107 hours on one dual-Xeon, dual-GPU node, for a mean rate of 23 million particles per second per step.  We find \textsc{Abacus} is in good agreement with \textsc{Ramses} and \textsc{Pkdgrav3} and less so with \textsc{Gadget3}.  We validate our choice of time step by halving the step size and find sub-percent differences in the power spectrum and 2PCF at nearly all measured scales, with $<0.3\%$ errors at $k<10\ \mathrm{Mpc}^{-1}h$.  On large scales, \textsc{Abacus} reproduces linear theory better than 0.01\%.  Simulation snapshots are available at \url{http://nbody.rc.fas.harvard.edu/public/S2016}.
\end{abstract}


\section{Introduction} \label{sec:intro}
Cosmological $N$-body simulations are the primary tool for forward modeling the theory of large-scale structure to observable quantities like the spatial distribution of galaxies.  As observations improve, the comparison of the forward model with observations becomes increasingly sensitive to systematic errors in the $N$-body simulations.  Some systematics can be checked analytically, such as the recovery of linear theory on large scales, but most rely on ``convergence testing'', in which a parameter of the simulation (such as the time step) is moved towards the continuum value until the answer stops changing (to some tolerance).  Such tests can be prohibitively expensive (see \citealt{DeRose+2018} for a recent exhaustive effort) and are not guaranteed to converge to the physical answer.

A common additional check is to compare the ``converged'' results from multiple, independent codes.  While not a guarantee of physical accuracy, agreement indicates control over systematics related to the numerics, to the extent that different codes use different numerical techniques.  This is the approach of code comparison projects like \cite{Heitmann+2008} and \citet[hereafter S2016]{S2016}.  The latter presents the code comparison project from the Euclid Cosmological Simulations Working Group, which compared the matter power spectrum from the \pkdgrav\citep{Potter+2017}, \ramses\citep{Teyssier_2001}, and \gadget\citep{Springel_2005} codes.

A third path to assessing code accuracy in the non-linear regime is through scale-free simulations.  In these tests, a power-law power spectrum is used in an expanding $\Omega_M=1$ background, such that the clustering on small scales should be a rescaling of the clustering on large scales at a later time.  Any deviation from this self-similarity must be due to finite box size, finite particle mass, or inaccurate numerics.  The breakdown of this self-similarity can be used to identify halo mass resolution limits and other complex non-linear systematics; this will be our approach in an upcoming paper (Joyce et al., in prep.).

In this work, we contribute \Abacus's result to the \citetalias{S2016} code comparison project.  \Abacus is a GPU-accelerated code for cosmological $N$-body simulations; it offers excellent force accuracy and minimal integration errors of the particle trajectory due to the small global timestep used.  It also employs a compact spline softening kernel, which minimizes leakage of force softening to large scales.  \Abacus's speed allows us to perform convergence tests at scale; we do not need to sacrifice volume or mass resolution to complete the tests in a reasonable amount of time with modest computational resources.

The paper is organized as follows.  In Section \ref{sec:abacus}, we discuss the \Abacus code and performance in the context of the \citetalias{S2016} simulation.  In Section \ref{sec:comparison}, we compare the \Abacus matter-field clustering results with \ramses, \gadget, and \pkdgrav.  In Section \ref{sec:validation}, we show validation tests for various \Abacus code parameters.  We discuss our findings in Section \ref{sec:discussion}.

\section{\Abacus} \label{sec:abacus}
\subsection{Method}
\Abacus is a code for cosmological $N$-body simulations based on the compact near-field/far-field force split developed in \cite{Metchnik_2009}.  In this approach, the domain is decomposed into $K^3$ cubic cells, and every particle belongs to one cell.  Particles in cells separated by fewer than near-field radius $R$ cells (typically 2) interact via the near-field force, which we compute as a direct pairwise summation of $1/r^2$ forces (or some appropriately softened form).  Particles in cells more distant than $R$ interact via the far-field force, which is computed with a multipole method.  Thus, every pairwise interaction is only present in either the near field or the far field.

The simplicity of the near-field computation offers a substantial performance and accuracy opportunity.  Due to the compact force split, the near-field force is exact (up to machine precision); thus, any force inaccuracy must arise from the far-field.  To increase total force accuracy, one thus only needs to increase the far-field multipole order $p$.  The challenge then becomes offsetting the computational load of doing so.  This is where the GPU performance helps: we can decrease $K$ to shift work from the far field into the near field, balancing the performance for the choice of $p$.  Modern GPUs excel at the kind of work $N$-body requires: compute-dense kernels consisting of a few simple mathematical operations repeated many times on a small amount of data.  In \Abacus, this performance translates quite directly into increased accuracy: the faster the near field becomes, the smaller the optimal $K$ becomes, allowing us to increase $p$ at fixed wall-clock time.

We organize particles into ``slabs'' one cell wide.  Particles are processed in a ``slab pipeline'': we load a rolling window of slabs into memory, compute forces and update particles on the central slab, write out the trailing slab, and then load a slab at the leading edge.  Only the central slab can be processed at a given time because $R$ slabs must be present on either side to compute the near force.  For $R=2$, this means 5 slabs must be in memory.  In practice, we allow \Abacus to read ahead by a few slabs, so we typically have 7 slabs in memory.  For $K=693$ (which we use for the \citetalias{S2016} simulation), the rolling window is thus 1\% of the total volume.

This is a substantial opportunity: since not all slabs have to be in memory, we don't need a large computer cluster; we can instead use a single node and store the slabs on hard drives, reading and writing them in an ordered sweep.  The raw compute power can be provided by GPUs for the near-field force and fast vectorized CPU implementations for the far-field force.

Of course, particles separated by more than the pipeline window size must ``talk'' to each other, too.  A single time step of Abacus thus consists of two primary sub-steps: \singlestep and \convolution.  \singlestep is the slab pipeline step detailed above.  As part of this pipeline, the multipole moments of the particle positions are computed in every cell and written to disk.  After \singlestep has completed, we thus have $K$ slabs of the cell multipoles.  The \convolution convolves these slabs with a ``derivatives'' tensor to produce a Taylor series approximation to the force in every cell.  We dub these the ``Taylors''.  The derivatives tensor is so called because it uses the derivative of the gravitational potential from the multipole moments to produce the acceleration.  This tensor is fixed for a given $K$, $R$, and multipole order $p$ and is pre-computed in a small amount of time.

The convolution is performed in Fourier space as a multiplication, so in detail we perform the YZ-FFT during \singlestep while we have the whole slab in memory (a slab spans all Y \& Z for a single X).  Thus, the convolution's task is to do the cross-slab X-FFT, apply the derivatives tensor in Fourier space, and do the inverse X-FFT to produce the Taylors.  The inverse YZ-FFT is done while applying the Taylors to a slab in \singlestep.

In this version of \Abacus, we compute the near field with brute-force $N^2$ summation.  Future versions of \Abacus will employ trees to accelerate the force computation in dense clusters.  Our choice of when to use a tree (and what leaf opening criteria to employ) will be determined by the usual efficiency-accuracy trade-off.

Although \Abacus largely employs single precision (32-bit floats) for particle kinematic data, positions are stored as offsets relative to cell centers.  This gains us an extra 9-10 bits of mantissa beyond the nominal 23 in IEEE 754.  Multipole and Taylor data is stored as 32-bit floats, but all internal computations are performed in double precision to avoid buildup of numerical imprecision.

Multipole order $p=8$ is our usual choice that balances performance and accuracy.  One way we test our accuracy is with the ``Ewald test'', in which we compute the forces on a random distribution of 65 K particles with a brute-force Ewald summation \citep{Ewald_1921} in quad-double precision.  Comparing with \Abacus's forces, we find the 99\% and median fractional errors are \SI{1.6e-4} and \SI{1.2e-5}, respectively.  We also use a ``homogeneous lattice'' test, in which a uniform grid of particles is set up such that the forces should be zero everywhere.  For $p=8$, the maximum deviation is \SI{2.6e-5}, in units of the displacement that would produce that force under the Zel'dovich Approximation \citep{Zeldovich_1970}, expressed as a fraction of the inter-particle spacing.

One reason we choose the multipole order to give such high force accuracy is that our domain decomposition is a structured mesh.  When computing forces on such a repeating structure, the force error patterns are likely to not be homogeneous and random; they will vary based on position in the cell and approximately repeat in every cell.  Such a spatially repeating error could readily appear in the power spectrum, which is one of the primary quantities we wish to measure from these simulations.

The primary \Abacus code paper is in preparation (Garrison, et al.).  A paper is also in preparation detailing our far-field method.  \Abacus has already been validated in a number of contexts, in internal convergence tests and against analytic theory \citep{Garrison+2016} and against other codes and emulators \citep{Garrison+2018}.

\subsection{Performance: Design} \label{sec:perf_design}
We present the performance of \Abacus for the \citetalias{S2016} $2048^3$ simulation on one node.  The performance and low memory requirements enabled by the exact force split mean that \Abacus does not need a computer cluster to complete large simulations in a reasonable amount of wall-clock time; indeed, \Abacus presently only supports single-node operation.

We built the node used in this work, called \hal, specifically for \Abacus using commodity computer hardware. \hal is a dual-socket Intel platform with two 14-core Intel Xeon Gold 6132 @ 2.60 GHz, 128 GB DDR4-2666 RAM, and two NVIDIA GeForce GTX 1080 Ti GPUs.  HyperThreading is disabled and the CPU frequency scaling governor is set to \texttt{performance}.  \hal is equipped with four 1 TB Samsung NVMe SSDs (two 970 Pro, and two 960 Pro).  We used the Intel compiler \texttt{icc} 17, and NVIDIA CUDA 9.2.  \hal cost about \$13000 and consumes approximately 1 kW under load.

The NVMe drives store the particle and convolution data.  Since we only hold 1\% of slabs in memory at a time, and since the CPU and GPU compute rate is so high, the drives holding the particles must be similarly fast.  For this work, we used the two 970 Pros for the particle data (440 GB), and the two 960 Pros for the convolution data (multipoles and Taylors, 102 GB total).  While we could have fit the whole $2048^3$ problem on a single SSD, we rely on multiple SSDs to provide the throughput to keep up with the GPU and CPU.  For larger simulations, we employ RAID arrays of HDDs, which offer lower performance but a better price/GB ratio.

During \singlestep, since we can compute the near field on GPUs and the far field on CPUs, we can overlap their computation.  Typically, a few CPU cores are dedicated to GPU communication, a few are dedicated to IO, and the rest are used for far-field forces and other CPU work.  Thus, we have the IO, GPU, and CPU operating in parallel.  For this work, we used 6 cores to prepare GPU work, 1 core for IO, and 21 for CPU work.

We carefully control the assignment of threads to cores (both OpenMP threads and our own GPU and IO threads).  This is to prevent threads from switching cores and interfering with each other and to maintain NUMA locality.  In most contexts, we statically schedule the OpenMP threads over $y$-rows, so particles will largely stay on their own NUMA node.

For this simulation, we use $K=693$, multipole order $p=8$, and near-field radius $R=2$.  This yields 25.8 particles per cell.  $K$ was chosen \textit{post hoc} to optimize the trade-off between the late-time $N^2$ work and the increased CPU work and IO for all time steps.

We note that the \citetalias{S2016} particle mass (\SI{1.2e9}{\hMsun}) is smaller than would be optimal for the current version of \Abacus on the \hal hardware.  At late times, the $N^2$ work from the largest halos dominates the runtime; the largest cell has over 200,000 particles by the end of the simulation.  To counteract this, we were forced to choose a relatively large $K$, which tends to under-fill the AVX-512 vectors in most cells and increase the FFT work.  This decreases far-field performance.  Furthermore, the multipole/Taylor data volume increases as $K^3$, thus slowing down the convolution step (which is strongly IO limited).  Even a factor of 2--3 increase in particle mass would cause the GPU work to be subdominant to the CPU work, increasing the total speed of the simulation.

The small force softening also leads to larger particle accelerations in the centers of halos.  Since \Abacus is a globally time-stepped code, this forces us to take one or two thousand time steps to $z=0$, instead of one or two hundred, as is common in codes with adaptive time stepping.  Future versions of \Abacus will address this with on-the-fly identification of halos and refined time stepping within those halos; we call this scheme ``microstepping''.  This will allow us to take larger global time steps and have the benefit of increasing the compute load for every time we load the particles.  This should bring the compute performance back in line with the disk performance.  The former currently outstrips the latter, except at late times for the low particle masses considered here.

For truly massive simulations, we are developing a parallel version of \Abacus based on the existing slab decomposition suitable for parallelization over a few dozen nodes.

While it may seem that the \hal hardware is specialized (e.g.~the combination of fast local disk and several GPUs on a single node), we note a trend in supercomputing towards this ``fat node'' design.  For example, the Summit\footnote{\url{https://www.olcf.ornl.gov/summit/}} supercomputer (number 1 on the Top500\footnote{\url{https://www.top500.org/list/2018/11/}}) has 1.6 TB of NVMe SSD and six NVIDIA Volta V100 GPUs per node.  We expect Abacus to be well-suited to Summit and Summit-like architectures.

\begin{figure}[ht!]
\plotone{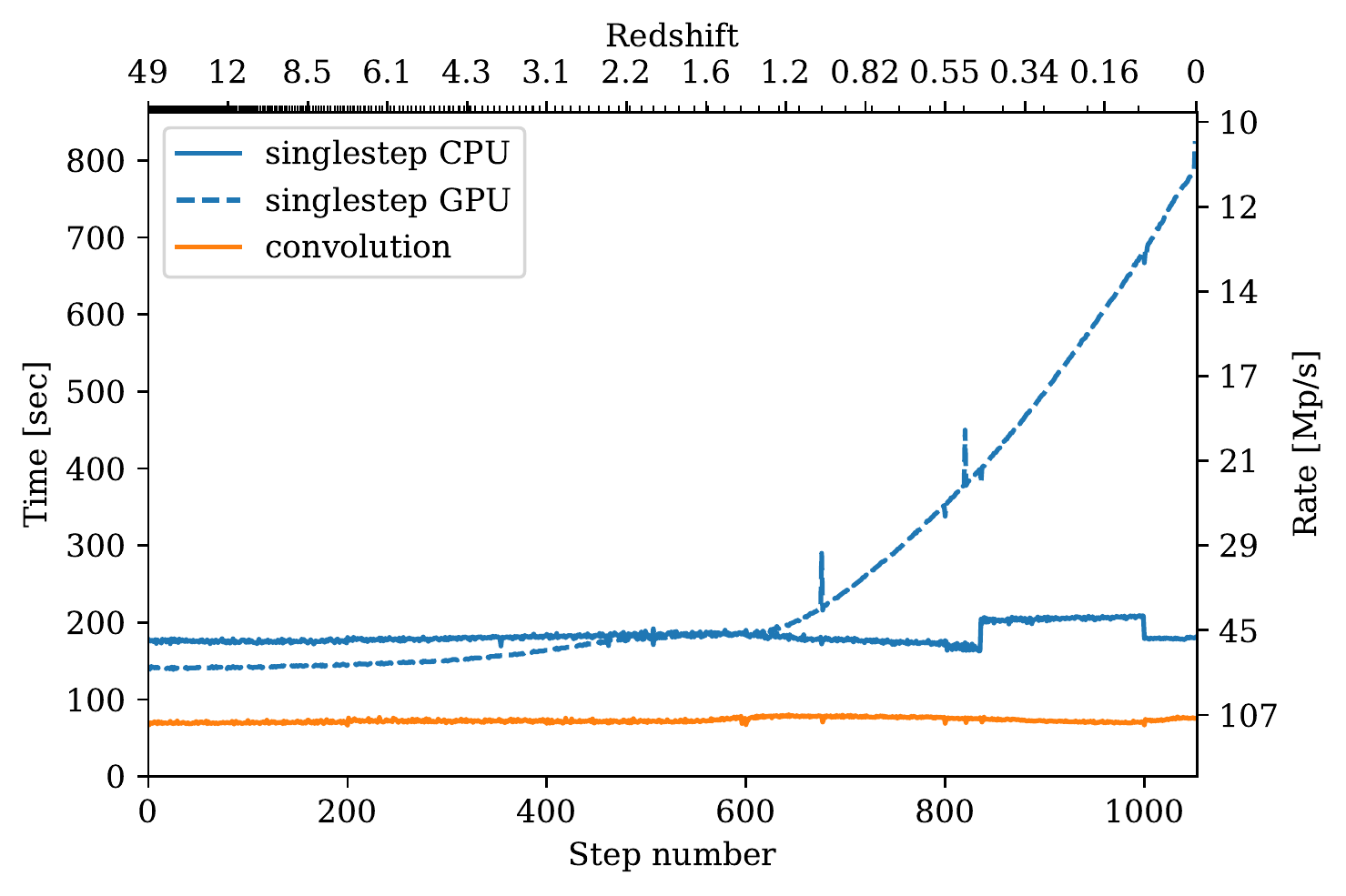}
\caption{\Abacus runtime per step. The \singlestep GPU and CPU work is overlapped, so the wall-clock time is the maximum of the two.  The convolution occurs as a separate step between \singlestep invocations.
The spikes in GPU runtime are the output steps, where the CPU is too busy to prepare work for the GPU at full speed.
Minor ticks on the redshift axis are steps of $\Delta z=0.1$.
\label{fig:timing}}
\end{figure}

\subsection{Performance: Results} \label{sec:perf_results}
\begin{figure}[]
\plotone{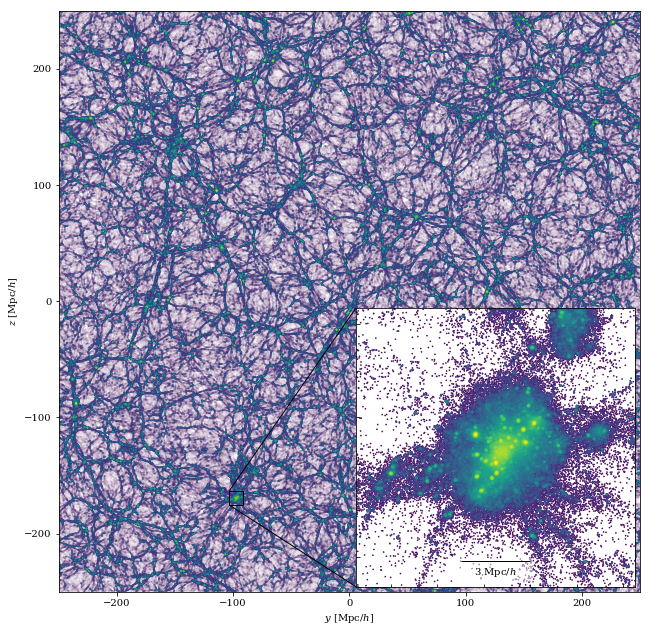}
\caption{A \SI{0.7}{\hMpc} thick slice through an \Abacus realization of the \citetalias{S2016} box at $z=0$.  Color indicates projected surface density.
\label{fig:pretty}}
\end{figure}

Using the exact particles provided by the Euclid Cosmological Simulations Working Group\footnote{\url{https://www.ics.uzh.ch/~aurel/euclid.htm}} [i.e.~no corrections to the initial conditions in the style of \citet{Garrison+2016}], \Abacus executed 1052 time steps from $z_\mathrm{init}=49$ to $z=0$ in 107 hours (4.5 days) for a mean rate of 23 million particles per second per step (Mp/s).  The \singlestep work took 87 hours, and the convolution 21 hours.  In \singlestep, the GPU work (near-field force) was fully masked by the CPU work until about redshift $z=1.5$ (step 600), after which it quickly became dominant (Fig.~\ref{fig:timing}).  \singlestep started at 46.0 Mp/s and ended at 10.8 Mp/s due to the increased GPU work.  Including the convolution, \Abacus started at 33.6 Mp/s and ended at 9.9 Mp/s.  See Fig.~\ref{fig:pretty} for a visualization of the final state.


A timing breakdown of the first \Abacus time step is given in Table \ref{tbl:timing}.  This timing is representative of all time steps, except for the increased GPU work towards late times, as noted in the table.

\begin{figure}[ht!]
\plotone{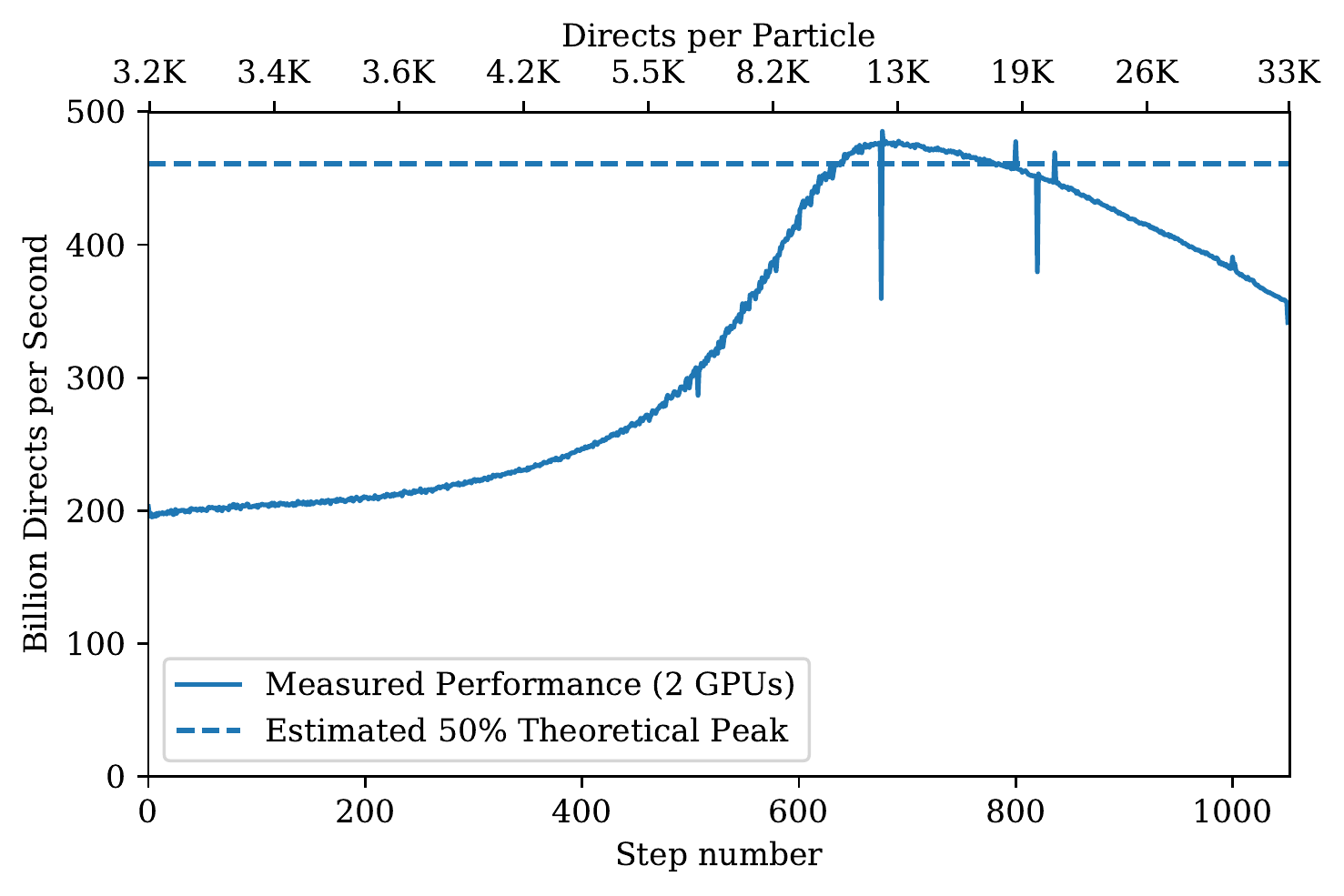}
\caption{GPU performance for the near-field pairwise force computation.  The theoretical maximum is computed assuming 10.6 TFLOPS per GPU and 23 FLOP per direct interaction (spline kernel).  The latter is a conservative lower limit and we surmise that our peak performance is actually larger than the 52\% of the ideal maximum shown here. \label{fig:gpu_performance}}
\end{figure}

In Fig.~\ref{fig:gpu_performance}, we show the measured GPU wall-clock performance in terms of number of pairwise spline interactions computed per second.  The performance increases as the compute density (interactions per particle) increases, peaking around step 700 ($z=1$) or \SI{1.2e4} interactions per particle.  Afterwards, the performance declines, possibly due to worsening load balancing from the increasing density contrasts between cells.

We also give a rough estimate of the theoretical peak performance of our two NVIDIA 1080 Ti GPUs.  We assume 10.6 TFLOPS per GPU (see above), which assumes all operations are fused multiply-add (FMA).  In our spline kernel, we count 22 additions and multiplies (not all of which are FMA), a reciprocal square root (\texttt{rsqrt)}, and a \texttt{min}.  We count the \texttt{rsqrt} as one FLOP and ignore the \texttt{min}, even though we expect these are poor approximations.  We thus derive a conservative 23 FLOP estimate, yielding a theoretical peak of 920 billion direct interactions per second (GDIPS).

We measure a peak \Abacus performance of 485 GDIPS, which is 52\% of our estimated theoretical peak.  We consider this excellent performance.  This measurement uses wall-clock time while at least one \Abacus GPU thread is running and thus includes PCIe bus transfer overheads and load imbalancing.

A ``notch'' of 10\% slower CPU performance is visible between steps 837 and 1000 in Fig.~\ref{fig:timing}.  After step 836, the simulation was manually paused for several minutes to run \texttt{fstrim} on the SSDs to ensure consistent write performance (we had observed catastrophic write performance decreases in the recent past that were fixed with TRIM).  Upon resuming, the \singlestep performance was slightly slower in memory-bandwidth-bound operations like Transpose Positions.  At step 1000, the simulation automatically paused for several minutes to write a backup state.  Upon resuming, the bandwidth issue disappeared.  We do not have compelling explanation for this issue, but it also had no impact on the wall-clock time, since it was completely masked by the GPU compute time.


\begin{deluxetable}{l|D|D|r|l}
\tablecaption{Wall clock timing for the first \Abacus time step of the \citetalias{S2016} box, $z=49$.  Units of ``Mp/s'' mean millions of particles per second.  ``DP-FLOPS'' means double-precision floating-point operations per second.  Only rates for the dominant sub-steps are shown.  Percentages are relative to their parent step.  ``Non-blocking'' means other CPU actions can proceed while that action is running. \label{tbl:timing}}
\tablehead{\colhead{Action} & \twocolhead{Time [s]} & \twocolhead{\%} & Rate & \colhead{Notes}}
\decimals
\startdata
Total & 255 & 100 \% & 33.6 Mp/s & 9.9 Mp/s at $z=0$ \\
\hline
\quad singlestep & 187 & 73 \% & 46 Mp/s \\
\qquad CPU Work & 187 & 100 \% & 46 Mp/s \\
\qqquad CUDA Initialization & 7.3 & 3.9\% & \ & Pinning memory \\
\qqquad Check Slab Integrity & 1.6 & 0.9\% & \\
\qqquad Transpose Positions & 3.0 & 1.6 \% & \\
\qqquad Prepare Near Force & 7.6 & 4.1 \% & \\
\qqquad Taylor Force & 85.2 & 45.5 \% & 100 Mp/s \\
\qqquad Kick & 10.1 & 5.4 \% & \\
\qqquad Drift & 7.8 & 4.2 \% & \\
\qqquad Multipoles & 48.3 & 25.8 \% & 177 Mp/s \\
\qqquad Finish & 14.7 & 7.9 \% & \\
\qqquad Waiting for GPU or IO & 1.0 & 0.5 \% & \\
\qquad GPU Near Force (non-blocking) & 137 & 73 \% & 62 Mp/s & 11 Mp/s at $z=0$ \\
\qquad Disk IO (all non-blocking) & . & . & \\
\qqquad Particle Data Read & 100 & 53 \% & 2.8 GB/s & 285 GB read \\
\qqquad Particle Data Write & 112 & 60 \% & 2.5 GB/s & 285 GB written \\
\qqquad Taylors Read & 30.6 & 16 \% & 1.7 GB/s & 108 GB read \\
\qqquad Multipole Write & 27.6 & 15 \% & 1.9 GB/s & 108 GB written \\
\hline
\quad convolution & 68 & 27 \% & \ & All work is CPU \\
\qquad Array Swizzle & 19 & 28 \% & \\
\qquad Convolution Arithmetic & 19 & 28 \% & \SI{5.8e9}{} & DP-FLOPS/core\\
\qquad z-FFT & 10 & 15 \% &  \\
\qquad Inverse z-FFT & 10 & 15 \% & \\
\qquad Wait for IO & 10 & 15 \% & \\
\enddata
\end{deluxetable}

\section{Code Comparison Results} \label{sec:comparison}
\subsection{Power Spectrum}
\begin{figure}[hp]
\plotone{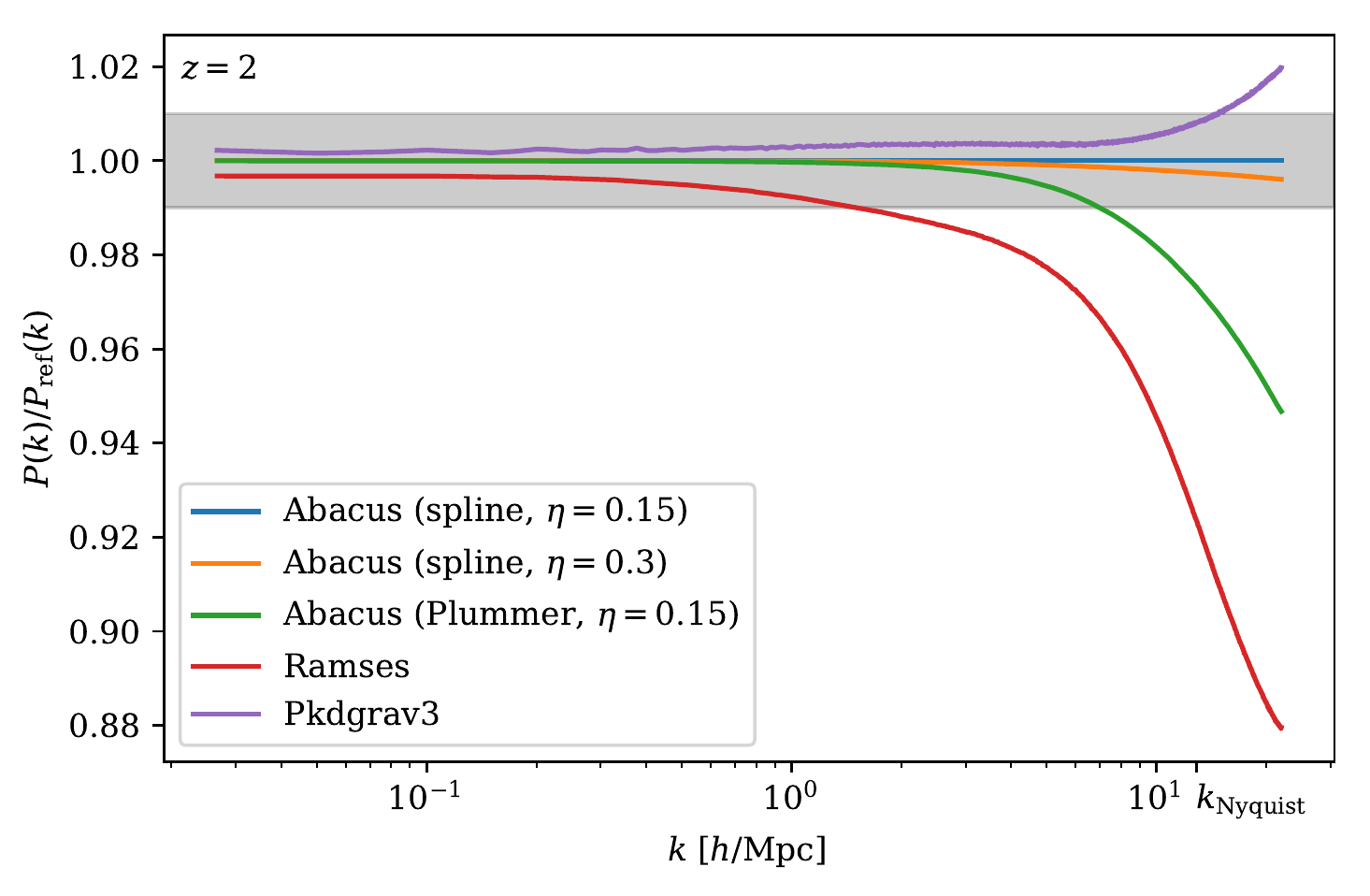}
\caption{Comparison of power spectra at $z=2$.  \gadget particles were not available for this redshift. \label{fig:power_z2}}
\end{figure}

\begin{figure}[hp]
\plotone{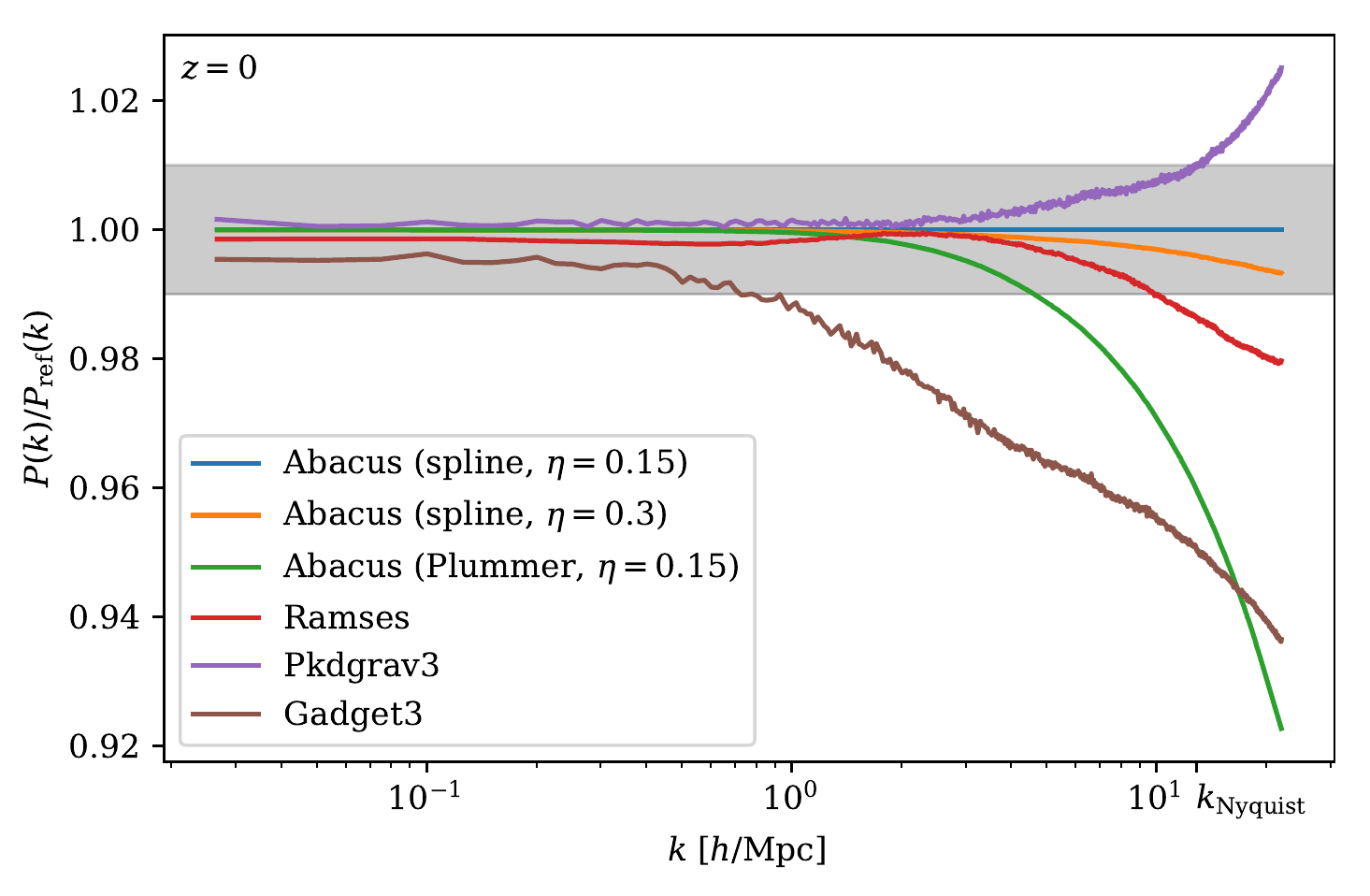}
\caption{Same as Fig.~\ref{fig:power_z2} (comparison of power spectra) but at $z=0$. \label{fig:power_z0}}
\end{figure}

We repeat the power spectra tests of \citetalias{S2016} on the $z=0$ and $z=2$ particles (except for \gadget $z=2$ particles, which were unavailable) and add \Abacus's results.  We use our own power spectrum code, which uses triangle-shaped (TSC) cloud mass assignment and TSC-alias window deconvolution \citep{Jing_2005}.  We use a $3500^3$ FFT mesh and find excellent agreement with the previously reported results.  The \Abacus result lies between \ramses and \pkdgrav at both $z=2$ \& $z=0$ (Figures \ref{fig:power_z2} \& \ref{fig:power_z0}).

As observed in \citetalias{S2016}, the codes do not agree even on the largest scales at both redshifts at the 0.5\% level.  This motivates us to check the analytic linear theory prediction of the power spectrum on these scales in Section \ref{sec:linear_theory}.  The largest disagreements are on the smallest scales, however.  This motivates our checks of the effects of time step and softening in Sections \ref{sec:timestep} \& \ref{sec:softening}.

\subsection{Two-Point Correlation Function}
We extend the analysis of \citetalias{S2016} to the small-scale two-point correlation function (2PCF).  We use the \textsc{corrfunc} code \citep{Corrfunc} to measure the auto-correlation of the matter density field out to \SI{1}{\hMpc}.  We first downsample the particles by a factor of two to reduce the pair-counting runtime.

\begin{figure}[ht]
\plotone{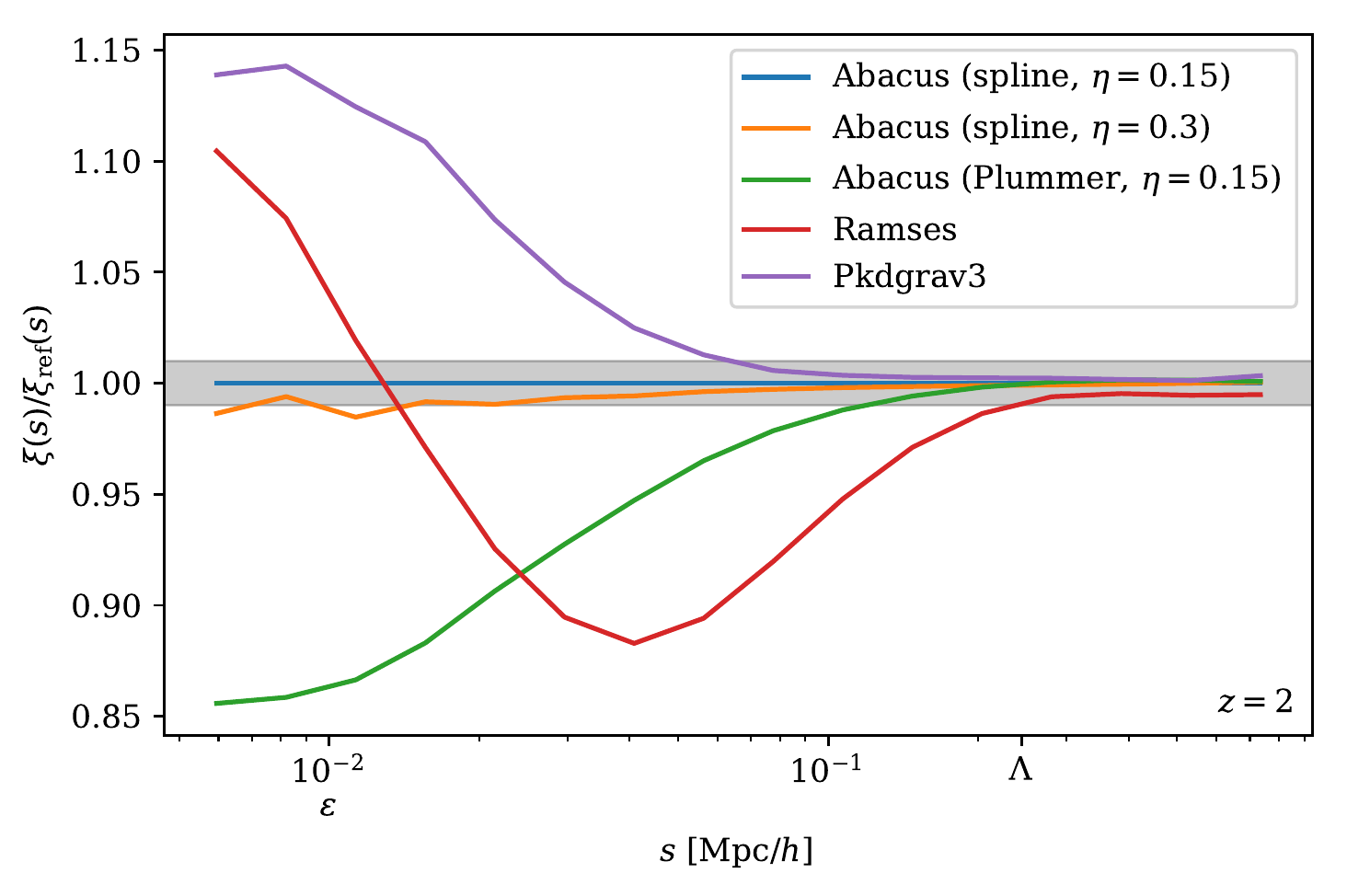}
\caption{Comparison of 2PCF at $z=2$.  $\epsilon$ marks the softening length, and $\Lambda$ marks the mean particle spacing.  Only compressed outputs were retained for \Abacus particles at this redshift, so some noise is apparent on small scales.  A larger binning was chosen to mitigate this effect. \gadget particles were not available for this redshift. \label{fig:2pcf_z2}}
\end{figure}

\begin{figure}[ht]
\plotone{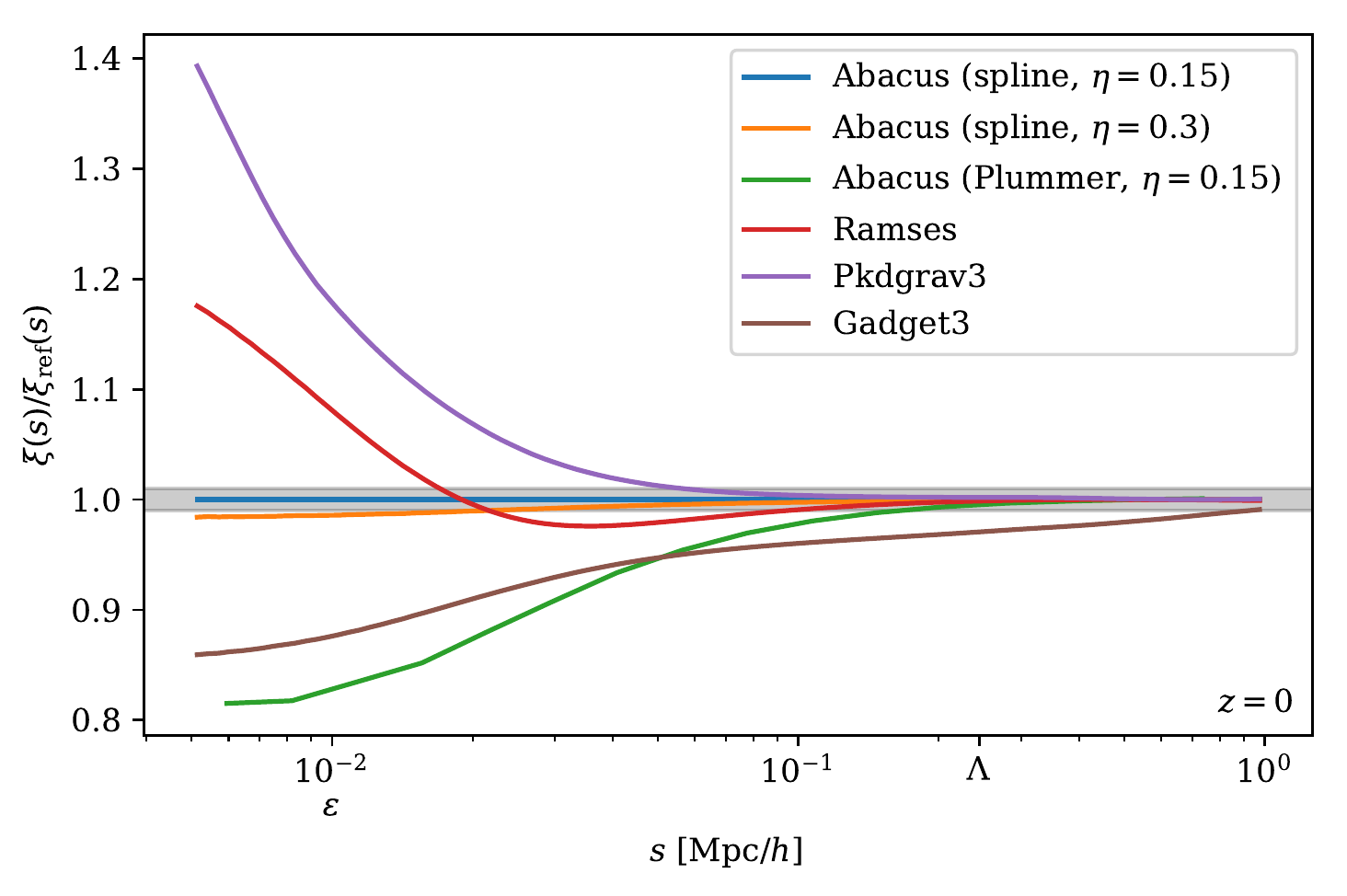}
\caption{Same as Fig.~\ref{fig:2pcf_z2} (comparison of 2PCF), but at $z=0$.  Full-precision outputs \Abacus outputs were available for all but the Plummer softening line, hence the coarser binning in that case. \label{fig:2pcf_z0}}
\end{figure}

The same trends that are visible in the small-scale power spectrum are visible in the 2PCF analysis (Figures \ref{fig:2pcf_z2} \& \ref{fig:2pcf_z0}).  The main trend that is qualitatively different from the power spectrum analysis is that the \ramses clustering amplitude exceeds that of \Abacus on the smallest scales.  This may be related to differences in the softening model.

\section{Validation} \label{sec:validation}
In addition to the end-to-end, standalone tests described in Section \ref{sec:abacus} (the Ewald and homogenous lattice tests), we validate the accuracy of Abacus specifically in the context of the \citetalias{S2016} simulation.  We test the recovery of linear theory, time step parameters, force softening model, and far-field and near-field accuracy.

\subsection{Linear Theory} \label{sec:linear_theory}
The \citetalias{S2016} codes do not agree on the power spectrum on the largest scales in the simulation.  Most notably, \ramses produces a 0.35\% power deficit compared to \Abacus at $z=2$, and \gadget produces a 0.45\% deficit at $z=0$.  Motivated by this failure to agree on linear theory, we test \Abacus's ability to recover the analytic result in the strongly linear regime.  We set up a $1024^3$ particle simulation with a $\sigma_8 = 0.817/200$ at $z=0$ and otherwise the same parameters as the \citetalias{S2016} simulation.  In particular, we hold fixed parameters that could plausibly affect the accuracy, like the particles per cell and the multipole order.  To best mimic \citetalias{S2016}, we used the ordinary Zel'dovich Approximation [i.e.~no corrections following \cite{Garrison+2016}].

\begin{figure}[ht]
\plotone{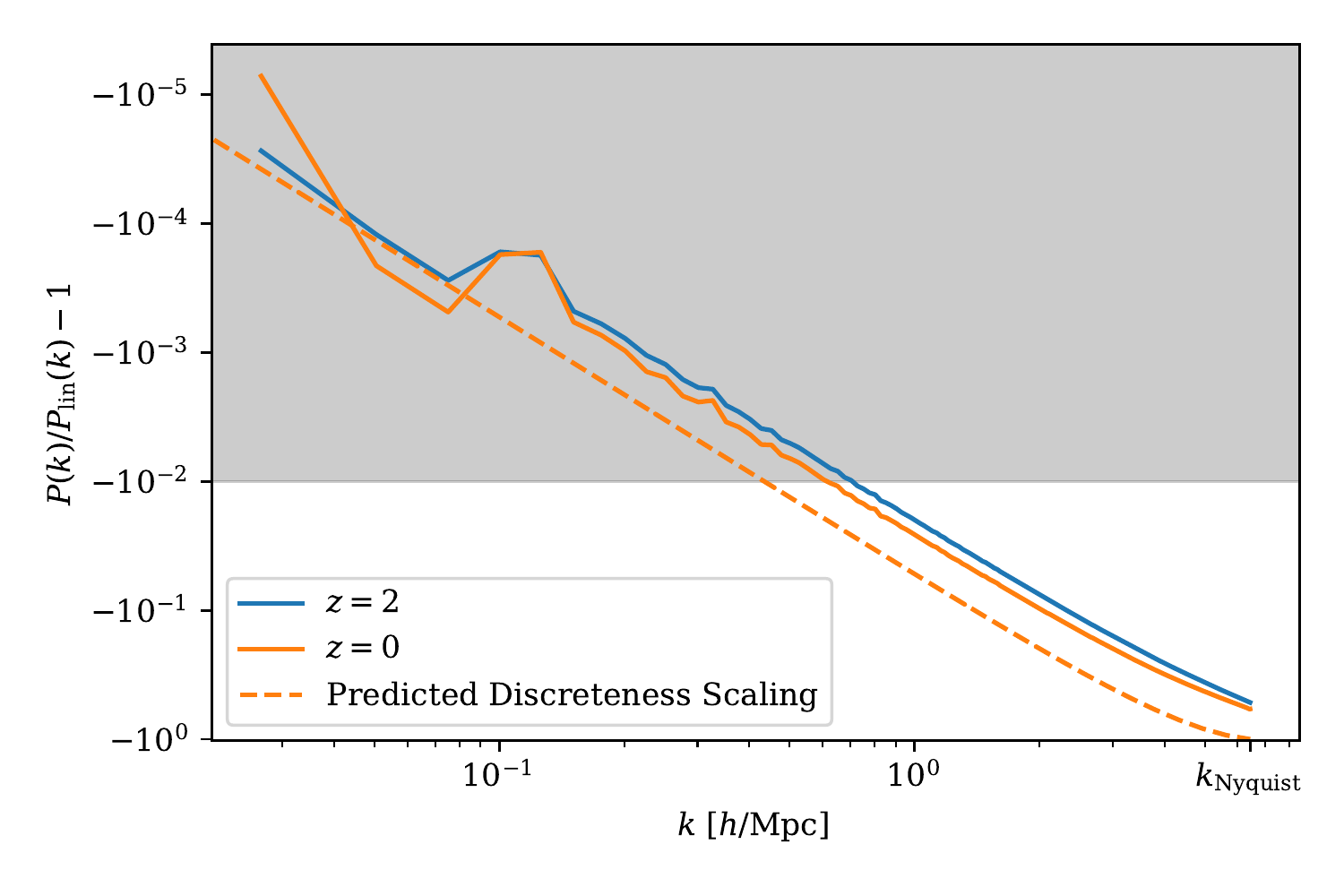}
\caption{Test of evolution the deeply linear regime ($\sigma_8 = 0.817/200$ at $z=0$).  \Abacus executed 137 time steps from $z_\mathrm{init}=49$ to $z=0$ and recovers the analytic linear theory prediction with excellent accuracy. The dashed line shows the predicted scaling of the suppression of growth rate due to discreteness, or finite particle mass, on the power spectrum. \label{fig:linear_theory}}
\end{figure}

We evolve the simulation from $z_\mathrm{init} = 49$ to $z=0$ using 137 time steps and compare the power spectrum at $z=2$ and $z=0$ to the linear power spectrum at those redshifts.  In both cases, we find better than 0.01\% agreement on the largest scales (Fig.~\ref{fig:linear_theory}).  We find a deficit of power on smaller scales, towards \kny.  This is expected.  An $N$-body system with finite particle mass should see a suppression of linear growth rate towards the Nyquist wavenumber of the particle sampling, independent of force softening or integration errors \citep{Marcos+2006,Garrison+2016}.  The wavenumber scaling of this effect is an excellent match to the analytic prediction (dashed line, Fig.~\ref{fig:linear_theory}), using the $k \ll \kny$ approximation of \cite{Marcos+2006}.  We see that $z=2$ consistently shows less suppression of power, as expected: the suppressed quantity is the growth \textit{rate}, so a higher redshift gives less time for the suppression to accumulate in an absolute sense.



\subsection{Time Stepping} \label{sec:timestep}
\subsubsection{Parametrization}
\Abacus is presently a globally-stepped code; all particles share the same time step.  At the beginning of each step, the time step $\Delta a$ is chosen based on three criteria:
\begin{enumerate}
\item the step size in $\Delta \log(a)$ units must not exceed \texttt{TimeStepDlna};
\item the step size must be less than \texttt{TimeStepAccel} times the maximum of $v_\mathrm{rms}/a_\mathrm{max}$ computed within each cell;
\item the step size times the maximum velocity must be less than 80\% of a cell width.
\end{enumerate}

The first criterion, set by \texttt{TimeStepDlna}, usually limits the step size at the beginning of the simulation before particle accelerations become large.  It ensures integration accuracy even in the linear regime.  We use a value of 0.03, or about 33 steps per $e$-fold of scale factor.  The successful linear theory test in Section \ref{sec:linear_theory} uses the same value and thus validates this choice.

The second criterion, controlled by \texttt{TimeStepAccel} (also called $\eta$), becomes the limiting factor as soon as two particles anywhere in the simulation come into a close orbit.  We nominally use a value of $\eta = 0.15$; we also try a value of $\eta = 0.3$ in Section \ref{sec:validation}, since we expect that our nominal choice is extremely conservative, given that we take 2200 global steps to $z=0$ as a result.

In detail, for the second criterion we also compute a global $v_\mathrm{rms}/a_\mathrm{max}$ and use the maximum of the global and cell-based values.  This protects us from taking catastrophically small time steps as a result of abnormally cold cells.

The third criterion simply ensures that particles drift by at most one slab per time step.  This is necessary in order to keep the rolling window of slabs in memory small.  In practice, we rarely trigger this criterion.

A large number of global steps is undesirable because (1) it increases the IO load, and (2) it wastes a large amount of computational effort integrating motions of low-acceleration particles.  This problem is particularly noticeable in the \citetalias{S2016} simulation, which has a small softening length and thus a high contrast in the dynamical time scale of halo and void particles.  We intend to address this in a future version of \Abacus with our ``microstepping'' scheme (see Section \ref{sec:perf_design}).
A large number of global steps does have the benefit of minimizing integration errors in the particle dynamics, however, which is useful in the context of checking code convergence.

\subsubsection{Validation}
We investigate the effect of varying the time-stepping parameter $\eta$, or \texttt{TimeStepAccel}, on the matter field power spectrum and 2PCF.  We try both $\eta = 0.15$ and $\eta = 0.3$.  \Abacus takes 2206 and 1052 time steps, respectively, to $z=0$.  We find sub-percent differences in the power spectrum to the smallest scales we measure ($k = \SI{22}{\perhMpc}$); the differences fall to 0.4\% above $\kny = \pi/\Lambda = \SI{12.9}{\perhMpc}$, where $\Lambda = L/N^{1/3}$ is the mean interparticle spacing.  The differences are even smaller at $z=2$.

Similarly, we find very small differences of less than 1.6\% across the whole measured range of the 2PCF at $z=0$, which extends down to $\epsilon/2$, or \SI{5}{\hkpc}.  This decreases to 1\% at $2\epsilon$.  Again, the differences are even smaller at $z=2$.

In both metrics, the error caused by increasing the time step to $\eta = 0.3$ is smaller by about an order of magnitude than the disagreements among the different codes.  Thus, we consider our choice of $\eta = 0.3$ to be sufficiently accurate.

\subsection{Softening} \label{sec:softening}
\subsubsection{Parametrization}
We investigate the effect of different force softening laws on our results.  Our nominal results use spline softening, but we also perform a simulation with Plummer softening.  In both cases, we use a Plummer-equivalent comoving softening length of \SI{10}{\hkpc} (see below). We use a timestep parameter of $\eta = 0.15$, which is the finer of the two timestep criteria investigated above.

In Plummer softening \citep{Plummer_1911}, the $\bfF(\bfr) = \bfr/r^3$ force law is modified as
\begin{equation}\label{eqn:plummer}
\bfF(\bfr) = \frac{\bfr}{(r^2 + \epsilon_p^2)^{3/2}},
\end{equation}
where $\epsilon_p$ is the softening length.  This softening is very fast to compute but is not compact, meaning it never explicitly switches to the exact $r^{-2}$ form at any radius (in contrast with spline softening). This affects the growth of structure on scales much larger than $\epsilon_p$, as we will see below.

Spline softening is an alternative in which the force law is softened for small radii but explicitly changes to the unsoftened form at large radii.  Traditional spline implementations split the force law into three or more piecewise segments \citep[e.g.~the cubic spline of][]{Hernquist_Katz_1989}; we split only once for computational efficiency and to avoid code path branching\footnote{We implement our split as a single \texttt{min} operation which compiles to a conditional move rather than a costly conditional jump.}.  We derive our spline implementation by considering a Taylor expansion in $r$ of Plummer softening (Eq.~\ref{eqn:plummer}) and requiring a smooth transition at the softening scale up to the second derivative\footnote{A Taylor expansion in $r^2$ is also possible, but we discard that solution due to a large plateau of constant angular frequency near $r\sim 0$ that we worry might excite dynamical instabilities.}.  This gives
\begin{equation}
\bfF(\bfr) =
\begin{cases}
\left[10 - 15(r/\epsilon_s) + 6(r/\epsilon_s)^2\right]\bfr/\epsilon_s^3, & r < \epsilon_s; \\
\bfr/r^3, & r \ge \epsilon_s.
\end{cases}
\end{equation}
This was first presented in \cite{Garrison+2016}.

The softening scales $\epsilon_s$ and $\epsilon_p$ imply different minimum dynamical times (an important property, as this sets the  step size necessary to resolve orbits).  We always choose the softening length as if it were a Plummer softening and then internally convert to a softening length that gives the same minimum pairwise dynamical time for the chosen softening method.  For our spline, the conversion is $\epsilon_s = 2.16\epsilon_p$.

\subsubsection{Comparison}
In Figures \ref{fig:power_z2} \& \ref{fig:power_z0}, we see that Plummer softening produces a significant suppression of small-scale power.  The range is notable too: 1\% effects extend even to $k$ below $\kny$, which itself is 24 times larger than the softening scale.

We see the same trend in the two-point correlation function in Figures \ref{fig:2pcf_z2} \& \ref{fig:2pcf_z0}: the suppression of clustering extends to many times the softening length.

In both the power spectrum and the 2PCF, using spline softening brings us into qualitatively better agreement with the other codes.  Due to its compact nature, we consider spline softening the more physically accurate of our two softening models.

\subsection{Far-field Force}
The main source of error in the far-field force is the finite multipole order $p$.  Our nominal value is $p=8$, which gives excellent force accuracy for near-field radius 2 (see Section \ref{sec:abacus}).  To quantify this in the context of \citetalias{S2016}, we re-ran the final state with $p=11$ and compared the far-field forces in the first 6 slabs to the $p=8$ result.  Measuring the fractional error as $|\mathbf{f_8} - \mathbf{f_{11}}|/|\mathbf{f_{11}}|$, we find the median error is \SI{6.8e-7}, with only 0.28\% of forces worse than \SI{1e-4}{}.

\subsection{Near-field Force}
The near-field force is essentially exact, since it is computed via brute-force $N^2$ summation (i.e.~no tree structures or other approximations are used).  The main source of uncertainty is use of single-precision (32-bit) floating point values for the positions and accelerations.  This mainly enters as round-off error in the accumulation of the accelerations, but there are other intermediate steps, like the re-centering of particle positions before they are sent to the GPU, that may suffer similarly.  To quantify these effects, we make a copy of the final state in double precision, evaluate the forces, and compare the forces to the single-precision answer.  We find that only 0.038\% of force errors are worse than \SI{1e-4}, which is an order of magnitude fewer than in the far-field.  The median fractional error is \SI{7.5e-7}{}.

\section{Discussion} \label{sec:discussion}
We have presented a realization of the \citetalias{S2016} code comparison simulation using our \Abacus $N$-body code.  \Abacus has excellent force accuracy properties that give us confidence that we are recovering the correct answer on most scales, especially in the linear regime where other codes disagree on the answer.  Indeed, our linear evolution tests show better than 0.01\% recovery of linear theory growth.  We have validated our time step and force accuracy parameters and found them to be conservative.

On small scales, the answer still depends on the choice of softening model.  Even matching dynamical times, we find that Plummer softening produces a significant suppression of small-scale power.  We consider this non-physical and prefer our compact spline softening, which brings our small-scale results closer to that of \ramses and \pkdgrav.  \gadget is still somewhat of an outlier, generally missing power across a broad range of scales.  Part of this could well have to do with softening model differences, given the large effect we saw when switching from Plummer to spline.  It may be illuminating to compare these results to those of other codes like \textsc{HACC} \citep{Habib+2013} and \textsc{2HOT} \citep{Warren_2013} to determine which differences arise from the force-solving technique and which arise from the softening model.

We have also demonstrated \Abacus's performance, which exceeds 30 million particles per second per step until $z=1.1$.  Afterwards, the near-field computation slows down due to the amount of clustering at this particle mass.  Overall, we achieve a mean rate of 23 Mp/s and measure GPU performance of over 50\% of the peak theoretical FLOPS.  Future enhancements to \Abacus will substantially increase our performance at this force and mass resolution.

\acknowledgments
We would like to thank Doug Ferrer and Marc Metchnik as co-authors of \Abacus, and Nina Maksimova for assistance in building \hal, the computer used to run these simulations.  We also would like to thank Aurel Schneider and Doug Potter for providing the \citetalias{S2016} particle snapshots, and the referee for helpful comments.  This work has been supported by grant AST-1313285 from the National Science Foundation and by grant DE-SC0013718 from the U.S. Department of Energy.  DJE is further supported as a Simons Foundation investigator.






\bibliography{references}

\end{document}